\documentstyle [12pt]{article}

\begin{document}
\begin{titlepage}

\begin{center}
{\Large\bf On the finite temperature $\lambda\varphi^{4}$ and 
Gross-Neveu models. Is there a first order phase transition in 
$(\lambda\varphi^{4})_{D=3}$ ?}\\
\vspace{.3in}{\large\em A.P.C.Malbouisson\footnotemark[1] and 
N.F.Svaiter\footnotemark[2]}\\
 Centro Brasileiro de Pesquisas Fisicas-CBPF\\ Rua Dr.Xavier
 Sigaud 150, Rio de Janeiro, RJ 22290-180 Brazil\\
\vspace{-1cm} 
\footnotetext[1]{Presently a post-doctoral fellow of CNPq (Brazil) 
at Ecolle 
Polytechnique (France) \\ \hspace*{0.53cm}e-mail: 
adolfo@orphee.polytechnique.fr}
\footnotetext[2]{e-mail:nfuxsvai@lca1.drp.cbpf.br}
\subsection*{\\Abstract}
\end{center}

We study the behavior of two diferent models at finite 
temperature in a $D$-dimensional spacetime. The first 
one is the $\lambda\varphi^{4}$ 
model and the second one  
is the Gross-Neveu model.  
Using the one-loop approximation we show that 
in the $\lambda\varphi^{4}$ model the thermal 
mass increase with the temperature while the thermal 
coupling constant decrese with the temperature.
Using this facts we establish that in
the $(\lambda\varphi^{4})_{D=3}$ model 
there is a temperature $\beta^{-1}_{\star}$ above which 
the system can develop 
a first order phase transition, where the 
origin corresponds to a metastable vacuum. 
In the massless Gross-Neveu model, we 
demonstrate that for $D=3$ the thermal correction 
to the coupling constant is zero. For $D\neq 3$ our
results are inconclusive.

\nopagebreak
Pacs numbers: 11.10.Ef, 11.10.Gh

\end{titlepage}
\newpage\baselineskip .37in
\section {Introduction}\

In the last years, there has been much interest in the 
nature of the electroweak phase transition. The 
high temperature effective potential in the 
standard and in the 
$(\lambda\varphi^{4})_{D=4}$ models 
have been calculated by many authors, where the 
contribution from multiloops diagrams has been 
taking into account. Several 
authors have pointed out the 
importance to known whether in
$(\lambda\varphi^{4})_{D=4}$ model the phase transition is of 
first or second order \cite{44}. Our 
interest in these issues was stimulated by some  
results of Ford and Svaiter 
concerning  the thermal dependence of the 
mass and coupling constant in 
$(\lambda\varphi^{4})_{D=4}$ model 
defined in a non-simple connected spacetime \cite{1}.
In the aforementioned paper these authors studied a 
neutral scalar field in a $D=4$ 
dimensional spacetime using the one-loop effective 
potential. The 
cases of trivial and non-trivial topology of the 
spacelike sections and 
finite temperature were discussed. The temperature 
and topological dependent renormalized mass 
and coupling 
constant were derived 
using the Speer and Bollini, Giambiagi 
and Domingues analytic
 regularization \cite{3} and a modified minimal 
subtraction renormalization procedure \cite{40}. In 
addition they have 
also discussed 
the possibility of vanishing the renormalized 
coupling constant in this model, 
as well as the limits of validity of the 
one-loop approximation. 
Some calculations studying such kind of problems 
was given recently by Elizalde and Kirsten and also Villareal \cite{5}. 
This last author improved the precedent results studying  
the 
two-loops corrections to the effective potential for  
scalar fields defined in a spacetime 
with non-trivial topology 
of the spacelike sections.

The two goals of this paper are the following. 
The first one is to extend 
the discussion of 
the massive self-interacting $\lambda\varphi^{4}$ 
model to an arbitrary 
$D$-dimensional spacetime, assuming trivial 
topology of the 
spacelike sections and to analize temperature 
effects in a model with 
asymptotic freedom. The second one is to discuss 
the existence of a first order 
phase transition in the massive 
$(\lambda\varphi^{4})_{D<4}$ 
model.

Besides Yang-Mills theories in $D=4$, 
the other known 
perturbative renormalizable 
asymptotically free theories with fermions are the 
Nambu-Jona-Lasinio and 
the Gross-Neveu models 
\cite{4}. In the latter, a $N$ component 
fermion field with a quartic 
self-interaction 
is assumed. The model is perturbatively   
renormalizable for $D=2$ and develops 
asymptotic freedom.

Working in a generic 
$D$-dimensional 
spacetime, we first calculate the one-loop corrections to 
the renormalized mass and coupling 
constant 
in the $\lambda\varphi^{4}$ model. 
We obtained that the thermal mass increase and the 
thermal coupling constant decrease with the temperature.
Still using 
the one-loop aproximation, the thermal 
correction to the renormalized coupling constant 
in the Gross-Neveu model
is obtained. We demonstrate that in 
the case $D=3$  the thermal 
correction to the coupling constant  
is zero. For $D\neq 3$ our results 
are inconclusive.

In many papers studying second order 
phase transition in the $\lambda\varphi^{4}$ model 
the temperature dependence 
of the coupling constant is neglected. This 
approach is reasonable since the 
variation of the mass with the temperature is the most 
important fact for a critical phenomena.
In this case, it is sufficient to consider the 
renormalized coupling constant 
as constant and the thermal mass drives the 
second order phase transition \cite{26}. In 
this paper we will examine the 
existence of a first order phase 
transition in the $\lambda\varphi^{4}$ 
model taking into account the thermal dependence 
of the coupling constant. Note that 
we will not deal with the 
system behavior in the neighborhood 
of a second order phase transition since we assume that 
the tree level mass squared $m^{2}$ 
is positive. This fact 
prevents the one-loop approximation to break down 
at low temperatures 
since there is no infrared divergences associated with vanishing 
masses. The result of our analysis can be summarized as 
follows: for $D<4$, there is a temperature 
$\beta^{-1}_{\star}$  
where the effective coupling constant 
vanishes. For temperatures $\beta^{-1}>\beta^{-1}_{\star}$, the 
renormalized coupling constant becomes 
negative and the system may suffer 
a first order phase transition. The 
effects of the radiative corrections 
is toward the direction of breaking a symmetry. 
Compare with  
the electroweak first order phase transition \cite{7}.  
We should note that at $\beta^{-1}=
\beta^{-1}_{\star}$ the 
system is still in an interacting phase. 
For $D<4$, there is  
a temperature 
where only the effective coupling constant 
$(\lambda(\beta)=\lambda-\lambda^{2}f(\beta))$ vanishes. 
All the higher
2n-points correlation functions 
do not vanish, therefore the model is 
not gaussian at the temperature 
$\beta_{\star}^{-1}$. This is an important point that was  
stressed by Weldon \cite{43}.

The study of the dependence 
of the coupling 
constant with the
temperature in QFT 
is well known in the literature. 
Many authors have studied 
such dependence in the 
$\lambda\varphi^{4}$ model and also in a abelian 
model like QED \cite{41}. Instead 
of using perturbative arguments, the use of the 
renormalization group equations 
allowed the investigation on the mass 
and coupling constant thermal 
dependence. Such program 
was implemented by Fujimoto, Ideura, 
Nakano and Yoneyama \cite{42}. These 
authors obtained results similar to ours 
in the $\lambda\varphi^{4}$ model. The behavior 
of the mass and coupling constant 
with the temperature are opposite, i.e. the 
renormalized mass increases if the 
system temperature increase as where the 
coupling constant
decreases. If we assume that the 
one-loop approximation provides 
trustable results we have the 
following situation: for 
temperatures 
above $\beta^{-1}_{\star}$ the 
renormalized coupling constant 
becomes negative. This 
behavior of the effective coupling 
constant is related to the fact 
that the model is non-asymptotically free. 
The growth of the renormalized 
coupling constant at large momenta is 
translated in our case to the 
temperature growth (in modulus) of 
this quantity. In $D=3$ for temperatures 
$\beta^{-1}>\beta^{-1}_{\star}$ the system 
develop 
a first order phase transition 
where the origin is a metastable vacuum.

In this paper we address only the 
one-loop approximation. It is not 
unreasonable to believe that our 
conclusions in the $\lambda\varphi^{4}$ 
model may be limited to this approximation. In fact, the 
behavior of the thermal correction to the 
coupling constant changes in the 
two-loops approximation. It was been 
shown by Funakubo and Sakamoto \cite{41}
that only for {\it low} 
temperatures the behavior of the 
thermal coupling constant remains 
the same as the obtained in the 
one-loop approximation. For 
high temperatures ($\beta^{-1}>>m$) the behavior is 
opposite i.e., the thermal correction 
is positive. Nevertheless this fact does not exclude the 
possibility of a first order 
phase transition at low temperatures in 
$(\lambda\varphi^{4})_{D=3}$. 
A more detailed discussion will appear in 
a forthcoming paper.

The paper is organized as follows. 
In section II we sketch 
the formalism of the effective potential. 
In section III, the massive 
self-interacting $\lambda\varphi^{4}$ model 
is analised.  
In section IV we repeat the calculations 
in the Gross-Neveu model. 
Conclusions are given in section V . 
In this paper 
we use $\frac{h}{2\pi}=c=1$.

\section{ The effective action and the effective potential at
 zero temperature.}

In this chapter we will review briefly the basic features of the 
effective potential associated with a
real massive self-interacting scalar field at 
zero temperature. 
Although the formalism of this 
section may be found in standard texbooks, we recall here its 
main results for 
completeness. Let us suppose a real 
massive scalar field $\varphi(x)$ with the usual 
$\lambda\varphi^{4}(x)$ self-interaction, defined in a static
spacetime. Since the manifold is static, there 
is a global timelike 
Killing vector field orthogonal to the 
spacelike sections. 
Due to this fact,
energy and thermal equilibrium have a precise meaning. For 
the sake of simplicity, 
let us suppose that the manifold is flat. In the path 
integral approach, the
basic object is the generating functional,
$$
Z[J]= < 0,out | 0,in >=
$$
\begin{equation}
\int{\cal D}[\varphi]\exp \{i[S[\varphi]+
\int d^{4}x J(x)\varphi(x)]\}
\end{equation}
where $ {\cal D}[\varphi]$ is the functional 
measure and $ S[\varphi]$ is the classical
action associated with the scalar field. The 
quantity $ Z[J] $ gives the
transition amplitude from the initial 
vacuum $ |0, in > $ to the final vacuum
$ |0,out > $ in the presence of some 
source $ J(x)$, which is zero outside
 some interval $ [-T,T] $ and inside this interval 
was switched adiabatically
 on and off. Since we are 
interested in the connected part of the time ordered products
 of the fields, 
we take the connected generating functional $W[J]$, 
as usual. This quantity is
defined in terms of the vacuum persistent amplitude by
\begin{equation}
e^{i W[J]}= <0, out|0, in >.
\end{equation}

The connected $n$-point function $ G_{c}(x_{1},x_{2},..,x_{n})$ 
is defined
by
\begin{equation}
G_{c}(x_{1},x_{2},..,x_{n})=\frac{\delta^{n} W[J]}{\delta J(x_{1})...
\delta J(x_{n})}|_{J=0}.
\end{equation}

Expanding $ W[J]$ in a functional Taylor series, 
the n-order coefficient of this
series will be the sum of all connected Feynman 
diagrams with $n$ external
legs, i.e. the connected Green's functions defined by eq.(3). Then
\begin{equation}
W[J]=\sum^{\infty}_{n=0}\frac{1}{n!}\int 
d^{4}x_{1}..d^{4}x_{n}~ G^{(n)}_{c}(x_{1},x_{2}..
..x_{n}) J(x_{1})J(x_{2})..J(x_{n}).
\end{equation}

The classical field $\varphi_{0}$ is given by 
the normalized vacuum 
expectation value of the field
\begin{equation}
\varphi_{0}(x)=\frac{\delta W}{\delta J(x)}=
\frac{<0, out|\varphi(x)|0, in >_{J}}{<0, out|0, in >_{J}},
\end{equation}
and the effective action $\Gamma[\varphi_{0}]$ is 
obtained by performing a 
functional Legendre transformation
\begin{equation}
\Gamma[\varphi_{0}]=W[J]-\int d^{4}x J(x)\varphi_{0}(x).
\end{equation}

Using the functional chain rule and the definition 
of $\varphi_{0}$ given
by eq.(5) we have
\begin{equation}
\frac{\delta\Gamma[\varphi_{0}]}{\delta\varphi_{0}}=-J(x).
\end{equation}

Just as $W[J]$ generates the connected Green's 
functions by means of a 
functional
Taylor expansion, the effective action can be 
represented as a functional 
power series around the value $\varphi_{0}=0$, 
where the coeficients are 
just the proper $n$-point
functions $ \Gamma^{(n)}(x_{1},x_{2},..,x_{n})$ i.e.,
\begin{equation}
\Gamma[\varphi_{0}]=
\sum^{\infty}_{n=0}\frac{1}{n!}\int d^{4}x_{1}d^{4}x_{2}..
.d^{4}x_{n}~\Gamma^{(n)}(x_{1},x_{2},..,x_{n})
~\varphi_{0}(x_{1})\varphi_{0}(x
_{2})..\varphi_{0}(x_{n}).
\end{equation}

The coefficients of the above functional expansion 
are the connected $1$
particle irreducible diagrams $(1PI)$. Actually, 
$\Gamma^{(n)}(x_{1},x_{2},..
.,x_{n})$ is the sum of all $ 1PI$ Feynman diagrams 
with $n$ external legs.
 Writing the effective action in powers of momentum 
(around the point where
all external momenta vanish) we have
\begin{equation}
\Gamma[\varphi_{0}]=\int d ^{4}x\biggl(-V(\varphi_{0})+
\frac{1}{2}(\partial_{\mu}
\varphi)^{2} Z[\varphi_{0}]~+~..\biggr).
\end{equation}

The term $ V(\varphi_{0})$ is called the effective 
potential\cite{8}\cite{9} .To express 
$ V(\varphi_{0})$ in terms of the $1PI$ 
Green's functions, let us write
$ \Gamma^{(n)}(x_{1},x_{2},..,x_{n})$ in 
the momentum space:
\begin{equation}
\Gamma^{(n)}(x_{1},x_{2},..,x_{n})=
\frac{1}{(2\pi)^{n}}
\int d^{4}k_{1}d^{4}k_{2}..d^{4}k_{n} (2\pi)^{4}
\delta(k_{1}+ k_{2} +..k_{n})\ e^{i(k_{1}x_{1}+..k_{n}x_{n})}
\tilde \Gamma^{(n)}(
x_{1},x_{2},..,x_{n}).
\end{equation}

Assuming that the model is translationally 
invariant, i.e. $\varphi_{0}$  
is constant over the manifold, we have
\begin{equation}
\Gamma[\varphi_{0}]=\int d^{4}x \sum^{\infty}_{n=1}
\frac{1}{n!}\biggr(\tilde\Gamma^{(n)}(0,0,..
.)(\varphi_{0})^{n}+...\biggr).
\end{equation}
If we compare eq.(9) with eq.(11) we obtain that
\begin{equation}
V(\varphi_{0})= -\sum_{n}\frac{1}{n!}\tilde\Gamma^{(n)}
(0,0,..)(\varphi_{0})^{n},
\end{equation}
then $ \frac{d^{n}V}{d\varphi^{n}_{0}} $ is the 
sum of the all $1PI$ diagrams
carring zero external momenta. Assuming that the 
fields are in equilibrium 
with a thermal reservoir at temperature 
$\beta^{-1}$, in the Euclidean 
time formalism, the effective potential 
$V(\beta,\varphi_{0})$ can be 
identified with the free energy density and 
can be calculated by 
imposing periodic (antiperiodic) boundary 
conditions on the bosonic (fermionic) fields. 

In the next section using the effective potential 
we will perform 
the one-loop renormalization 
of the $\lambda\varphi^{4}$ assuming that the system is  
in equilibrium with a 
thermal reservoir at temperature $\beta^{-1}$.
Since we are interested to 
make a paralel with the tricritical phenomena where in the 
tree level approximation with 
$V(\varphi)=m^{2}\varphi^{2}+
\lambda\varphi^{4}+\sigma\varphi^{6}$ predicts 
the existence of a first order phase transition if 
we allow the coefficient of 
the quartic term to be negative, we will 
evaluate the effective potential in a very unusual way. 
Instead of summing the series obtaining a log expression, and  
regularizing the model by introducing an 
ultraviolet cut-off in the Euclidean momenta, 
we prefer to use the principle 
of analytic extension in each 
term of the series. The advantage of this 
method lies in the fact that 
the dependence of mass and coupling constant 
with the temperature appear 
in a very straightforward way as well as the paralel with 
the tricritical phenomena.  

\section{ The one-loop effective potential in
the $\lambda\varphi^{4}$ model 
at zero and finite temperature.}\

Let 
us assume the following 
Lagrange density associated with a massive 
neutral scalar field:
\begin{equation}
{\cal L}= \frac{1}{2}(\partial_{\mu}
\varphi_{u})^{2}-\frac{1}{2}m^{2}_{0}~
\varphi_{u}^{2}-\frac{\lambda_{0}}{4!}\varphi^{4}_{u}\,
\end{equation}
where $\varphi_{u}(x)$ is the unrenormalized 
field and $m_{0}$ and $\lambda_{0}$
are the bare mass and bare coupling constant 
respectively. We may rewrite the 
Lagrange density as the usual form where the 
counterterms will appear explicity.
Defining the quantities
\begin{equation}
\varphi_{u}(x)= (1+\delta Z)^{\frac{1}{2}}\varphi(x)
\end{equation}
\begin{equation}
m^{2}_{0}=(m^{2}+\delta m^{2}) (1+\delta Z )^{-1}
\end{equation}
\begin{equation}
\lambda_{0}= (\lambda+\delta\lambda)(1+\delta Z)^{-2},
\end{equation}
and substituting eq.(14),(15) and (16) in eq.(13) we have
\begin{equation}
{\cal L}=\frac{1}{2}(\partial_{\mu}\varphi)^{2}-
\frac{1}{2}m^{2}\varphi^{2}-
\frac{\lambda}{4!}\varphi^{4}+\frac{1}{2}
\delta Z(\partial_{\mu}\varphi)^{2}-
\frac{1}{2}\delta m^{2}\varphi^{2}-
\frac{1}{4!}\delta\lambda\varphi^{4},
\end{equation}
where $\delta Z$, $\delta m^{2}$, and 
$\delta\lambda$ are the wave function, 
mass and coupling constant counterterms of the model. 
After the Wick rotation, in the one-loop 
aproximation, the effective potential is given by \cite{9}:
\begin{equation}
V(\varphi_{0})=V_{I}(\varphi_{0})+V_{II}(\varphi_{0})
\end{equation}
where,
\begin{equation}
V_{I}(\varphi_{0})= \frac{1}{2}m^{2}\varphi^{2}_{0}+
\frac{\lambda}{4!}
\varphi^{4}_{0}-\frac{1}{2}\delta m^{2}\varphi^{2}_{0}-
\frac{1}{4!}\delta\lambda\varphi^{4}_{0},
\end{equation}
and
\begin{equation}
V_{II}(\varphi_{0})=
\sum_{s=1}^{\infty}\frac{(-1)^{s+1}}{2s}\biggl(\frac{1}{2}
\lambda\varphi^{2}_{0}\biggr)
^{s}\int\frac{d^{D}q}{(2\pi)^{D}}
\frac{1}{(\omega^{2}+\vec{q}~^{2}+m^{2})^s}.
\end{equation}

Before continuing, we would like to 
discuss one important point. 
Performing analytic or dimensional 
regularization, we must 
introduce a mass parameter $\mu$, in 
terms of which dimensional 
analysis gives to the field a dimension $[\varphi]=\mu^{1/2(D-2)}$ 
and to the coupling constant a dimension 
$[\lambda]=\mu^{4-D}$. 
Mass has dimension of inverse length, i.e. $[\mu]=[m]=L^{-1}$, and 
the effective potential (the energy 
density per unit volume) 
has dimension of $L^{-D}$.

It is not difficult to extend the 
results given by eqs.(19) and 
(20) to finite temperature
states. After a Wick rotation, the functional integral 
runs over the fields that
satisfy periodic boundary conditions in Euclidean time. 
The effective action
can be defined as in the zero temperature case by a 
functional Legendre
transformation. Regularization and renormalization 
procedures follow
the same steps as in the zero temperature case. 
Although the counterterms 
introduced at finite temperature are the same as 
in the zero temperature 
case, the finite part of the physical parameters 
are temperature dependent. In this 
situation, since the sign of the thermal 
correction to the coupling constant 
is negative, the possibility of vanishing 
the renormalized coupling constant 
appears.

To study temperature effects we perform 
as usual the following 
replacement in the Euclidean region:
\begin{equation}
\int \frac{d\omega}{2\pi}\rightarrow\frac{1}{\beta}\sum_{n}
\end{equation}
and
\begin{equation}
\omega\rightarrow\frac{2\pi n}{\beta}
\end{equation}
where $\omega_{n}=\frac{2\pi n}{\beta}$ are 
the Matsubara frequencies.
Defining the dimensionless quantities:
\begin{equation}
c^{2}=\frac{m^{2}}{4\pi^{2}\mu^{2}}
\end{equation}
and
\begin{equation}
(\beta\mu)^{2}=a^{-1},
\end{equation}
 the Born terms plus one-loop terms 
contributing to the effective
potential give,
$$V(\beta,\varphi_{0})=V_{I}(\varphi_{0})+
V_{II}(\beta,\varphi_{0}) $$
where,
\begin{equation}
V_{I}(\beta,\varphi_{0})=\frac{1}{2}m^{2}\varphi^{2}_{0}
+\frac{\lambda}{4!}
\varphi^{4}_{0}-\frac{1}{2}
\delta m^{2}\varphi^{2}_{0}-\frac{1}{4!}
\delta\lambda\varphi^{4}_{0},
\end{equation}
and
\begin{equation}
V_{II}(\beta,\varphi_{0})=\frac{1}{\beta}
\sum^{\infty}_{s=1}\frac{(-1)^{s+1}}
{2s}\biggl(\frac{\lambda}{8\pi^{2}}\biggr)^{s}
\biggl(\frac{\varphi_{0}}{\mu}\biggr)^{2s}\int
\frac{d^{d}q}{(2\pi)^{d}}
A^{M^{2}}_{1}
(s,a).
\end{equation}
The function
\begin{equation}
A^{c^{2}}_{N}(s,a_{1},a_{2},..,a_{N})=
\sum^{\infty}_{n_{1},n_{2}..n_{N}=-\infty}
(a_{1}n_{1}^{2}+a_{2}n^{2}_{2}+...+a_{N}n^{2}_{N}+c^{2})^{-s}
\end{equation}
is the inhomogeneous Epstein zeta function\cite{10}, and finally
$$
M^{2}=\frac{1}{4\pi^{2}\mu^{2}}(\vec{q}~^{2})+c^{2}.
$$
Note that the mass parameter $\mu$ introduced in eqs.(23) and (24) 
will be used from now on, since we must have dimensionless functions 
when working with analytic extensions.

Let us define the modified inhomogeneous Epstein zeta function as
\begin{equation}
E^{c^{2}}_{N}(s,a_{1},a_{2},..a_{N})=
\sum^{\infty}_{n_{1},n_{2},..n_{N}=1}
(a_{1}n^{2}_{1}+..+a_{N}n^{2}_{N}+c^{2})^{-s}.
\end{equation}

Defining 
the new coupling constant and a new vacuum 
expectation value of the 
field $\phi$ (dimensionless for $D=4$),
\begin{equation}
g=\frac{\lambda}{8\pi^{2}}
\end{equation}
\begin{equation}
\frac{\varphi_{0}}{\mu}=\phi
\end{equation}
\begin{equation}
k^{i}=\frac{q^{i}}{2\pi\mu}
\end{equation}
we rewrite eq.(26) withouth use the definition 
of the inhomogeneous 
Epstein zeta function as,
\begin{equation}
V_{II}(\beta,\phi)=
\mu^{D}\sqrt{a}\sum^{\infty}_{s=1}
\frac{(-1)^{s+1}}{2s}g^{s}\phi^{2s}\sum^{\infty}_
{n=-\infty}\int d^{d}k\frac{1}{(an^{2}+c^{2}+\vec{k}~^{2})^{s}}.
\end{equation}
To regularize the model we will use a mix between 
dimensional and zeta 
function analytic regularizations. Let us first use dimensional 
regularization\cite{11}. Using the well known result,
\begin{equation}
\int\frac{d^{d}k}{(k^{2}+a^{2})^{s}}=
\frac{\pi^{\frac{d}{2}}}{\Gamma(s)}\Gamma(s-\frac{d}{2})
\frac{1}{a^{2s-d}},
\end{equation}
eq. (32) becomes
\begin{equation}
V_{II}(\beta,\phi)=
\mu^{D}\sqrt{a}\sum^{\infty}_{s=1}\frac{(-1)^{s+1}}{2s}
g^{s}\phi^{2s}\frac{\pi^{\frac{d}{2}}}
{\Gamma(s)}\Gamma(s-\frac{d}{2})
\sum^{\infty}_{n=-\infty}
\frac{1}{(an^{2}+c^{2})^{s-\frac{d}{2}}}.
\end{equation}
Defining,
\begin{equation}
f(D,s)=f(d+1,s)=\frac{(-1)^{s+1}}{2s}
\pi^{\frac{d}{2}}\Gamma(s-\frac{d}{2})
\frac{1}{\Gamma(s)}
\end{equation}
and substituting eqs.(27) and (35) in eq.(34) we obtain,
\begin{equation}
V_{II}(\beta,\phi)=\mu^{D}\sqrt{a}\sum^{\infty}_{s=1}f(D,s)g^{s}
\phi^{2s}A^{c^{2}}_{1}(s-\frac{d}{2},a).
\end{equation}

As we will soon see, the terms $s\leq\frac{D}{2}$ are 
divergent and we 
will regularize the one-loop effective potential using  
the Principle
of the Analytic Extension. Let us assume that 
each term in the series of
 the one-loop effective potential $ V(\beta,\phi)$ is the 
analytic extension of these terms, defining in the 
beginning in an open connected set. 
To render the discussion more general, let us discuss the 
process of the analytic continuation of the 
modified inhomogeneous Epstein
zeta function given by eq.(28). For $ Re(s) > \frac{N}{2}$, 
the $E^{c^{2}}
_{N}(s,a_{1},a_{2},..a_{N}) $ converges and 
represent an analytic function
of $ s$, so $Re(s) > \frac{N}{2} $ is the 
largest possible domain of the 
convergences of the series. This means that in eq.(36) 
in the case $D=4$ only the terms
$s=1$ and $s=2$ are divergent. The term $s=1$ 
is the divergent 
one-loop diagram 
of the connected two-point function and it 
contributes with a quadratic 
divergence. The $s=2$ term is the divergent 
one-loop diagram of the connected four point 
function, and it contributes to the effective 
potential with a logarithmic 
divergence. Using a Mellin transform it is 
possible to find the analytic 
extension of the modified inhomogeneous Epstein zeta function. 
After some calculations using Kirsten's results 
\cite{12}, we have:
\begin{equation}
V_{II}(\beta,\phi)=\mu^{D}\sum^{\infty}_{s=1}
f(D,s)g^{s}\phi^{2s}\sqrt{\pi}
\biggl(\frac{m}{2\pi\mu}\biggr)^{D-2s}\frac{1}
{\Gamma(s-\frac{d}{2})}
\biggl(\Gamma(s-\frac{D}{2})+4\sum^{\infty}_{n=1}
\biggl(\frac{mn\beta}{2}\biggr)^{s-\frac{D}{2}}
K_{\frac{D}{2}-s}(mn\beta)\biggr)
\end{equation}
where $K_{\mu}(z)$ is the Kelvin function \cite{13}.

It is not difficult to show that:
\begin{equation}
V_{II}(\beta,\phi)=
\mu^{D}\sum^{\infty}_{s=1}g^{s}\phi^{2s}h(D,s)
\biggl(\frac{1}{2^{\frac{D}{2}-s+2}}
\Gamma(s-\frac{D}{2})
(\frac{m}{\mu})^{D-2s}+\sum^{\infty}_{n=1}\biggl(\frac{m}
{\mu^{2}\beta n}\biggr)^{\frac{D}{2}-s}
K_{\frac{D}{2}-s}(mn\beta)\biggr)
\end{equation}
where:
\begin{equation} h(D,s)=\frac{1}{2^{\frac{D}{2}-s-1}}\frac{1}
{\pi^{\frac{D}{2}-2s}}\frac{(-1)^{s+1}}{s}
\frac{1}{\Gamma(s)}.
\end{equation}
If we  
suppose that $D=4$, the  
model is perturbatively renormalizable 
and an appropriate choice 
of $\delta m^{2}$ and $\delta \lambda$ will render the 
analytic extension of the terms of the 
series in $s$ in the effective 
potential analytic functions in the neighbourhood 
of the poles $s=1$ and $s=2$ respectively. 

The idea to extend the definition of 
an analytic function to a larger domain 
(analytic extension) and subtract poles 
was exploited by Speer, Bollini and 
others. In the method used by Bollini, 
Giambiagi and Domingues, a complex parameter 
$s$ was introduced as an expoent of the 
denominator of the loop expressions 
and the integrals are well defined analytic 
functions of the parameters in 
the region $Re(s)>s_{0}$ for some $s_{0}$. 
Performing an analytic extension of 
the expression for $Re(s)\leq s_{0}$, poles 
will appear in the analytic 
extension and the final expression becomes 
finite after a renormalization procedure. 
To find the exact form of the counterterms let us 
use the renormalization conditions
\begin{equation}
\frac{\partial^{2}}{\partial\phi^{2}}
V(\beta,\phi)|_{\phi=0}=m^{2}\mu^{2}
\end{equation}
and
\begin{equation}
\frac{\partial^{4}}
{\partial\phi^{4}}V(\beta,\phi)|_{\phi=0}=\lambda\mu^{4}.
\end{equation}
Since the vacuum expectation value of the field has 
been chosen to be constant,
there is no need for wave function renormalization. 
Substituting 
eqs.(25),(38) and (39) in eqs.(40) and (41) it 
is possible to find the 
exact form of the countertems in such a way that 
they cancel the polar 
parts of the analytic extension of the terms $s=1$ and $s=2$.
Note that we are using a "modified" minimal 
subtraction renormalization scheme where the mass 
and coupling constant  
counterterms are poles at the physical values of $s$.
It is straighforward to 
show that both $\delta m^{2}$ and $\delta\lambda$ are 
temperature independent.  
If a model at zero temperature is renormalizable with 
some counterterms it is 
also renormalizable at finite temperature 
with the same counterterms. 
This result was obtained in all orders of perturbation 
theory by Kislinger 
and Morley \cite{22}. In the 
neighbourhood of the poles $s=1$ and $s=2$, the regular 
part of the analytic  extension of inhomogeneous Epstein 
zeta function has two contributions: one which is 
temperature 
independent and that can be absorbed by the 
counterterms and 
another that is temperature dependent and cannot be 
absorbed by 
the counterterms. It is clear that the temperature 
dependent mass 
is proportional to the regular part of 
the analytic extension of 
the inhomogeneous Epstein zeta function in 
the neighborhood of 
the pole $s=1$. The same argument can be 
applied to the renormalized 
coupling constant. The thermal contribution 
to the renormalized 
coupling constant is proportional to the 
analytic extension of 
the inhomogeneous Epstein zeta function 
in the neighborhood of 
the pole $s=2$. The choice of the 
renormalization point $\phi=0$ 
implies that only the regular part in the 
neighborhood of the 
pole $s=1$ will appear in 
the renormalized mass. In the next 
section (where massless 
self-interacting fermion fields are 
studied) we will show  that all 
the terms of the series of 
the effective potential contribute 
to the renormalized mass 
and coupling constant and the sign 
of the thermal coupling 
constant cannnot be computed 
for $D\neq 3$. From the above 
discussion we can write 
\begin{equation}
-\tilde\Gamma^{(2)}(p=0,\beta,\lambda,m)=
m^{2}(\beta)=m^{2}+\Delta m^{2}(\beta)
\end{equation}
and
\begin{equation}
-\tilde\Gamma^{(4)}(p=0,\beta,\lambda,m)=
\lambda(\beta)=\lambda+\Delta\lambda(\beta),
\end{equation}
where $m^{2}(\beta)$ and $\lambda(\beta)$ are respectively 
the temperature 
dependent renormalized mass squared and 
coupling constant.  It can be 
directly shown that the thermal contribution to the 
renormalized mass squared is given by:
\begin{equation}
\Delta m^{2}(\beta)-
\Delta m^{2}(\infty)=\frac{1}{8\pi^{2}}\lambda
\sum^{\infty}_{n=1}\frac{m}{\beta n}K_{1}(mn\beta).
\end{equation}
Using the asymptotic representation of 
the Bessel function 
$K_{n}(z)$ for small arguments
$$ K_{n}(z)\cong\frac{1}{2}\Gamma(n)
(\frac{z}{2})^{-n}~~,z\rightarrow 0 ~~ n=1,2,..,$$
we obtain that at high temperatures the 
temperature dependent 
mass squared is proportional 
to $\lambda\beta^{-2}$ \cite{19}. 
The result given by eq.(44) was 
also obtained by Braden \cite{21} using  
Schwinger's proper time method. 
The same author also discussed the 
two-loop effective potential and 
the problem of overlapping divergences 
where the possibility of temperature
 dependent counterterms appears. 
Nevertheless these divergences must cancel as it was stressed by 
Kislinger and Morley \cite{22}.

Based uppon the same arguments 
previously used, the thermal 
contribution to the renormalized coupling constant is given by:
\begin{equation}
\Delta\lambda(\beta)-\Delta\lambda(\infty)=-\frac{3}{8\pi^{2}}
\lambda^{2}\sum^{\infty}_{n=1}K_{0}(mn\beta).
\end{equation}
The Bessel function $K_{0}(z)$ is 
positive and decreases for $z>0$. 
Therefore let us present an 
interesting result: the renormalized 
coupling constant attains its 
maximum at zero temperature 
$(\beta^{-1}=\infty)$ and 
decreases monotonically as the 
temperature increases. In other words, 
the thermal contribution 
to the renormalized coupling constant 
$\Delta\lambda(\beta)-\Delta\lambda(\infty)$ is negative, and 
increases in modulus with the temperature. 
The same result was 
obtained by Fujimoto, Ideura,
Nakano and Yoneyama using the 
renormalization group equations 
at finite temperature \cite{42}. 
Once we are discussing 
thermal effects, in the limit of zero temperature  the thermal 
contribution to the mass and coupling 
constant must vanish 
$(\tilde\Gamma^{(2)}(p=0,\beta=\infty,\lambda,m)=-m^{2}$ and 
$(\tilde\Gamma^{(4)}(p=0,\beta=\infty,\lambda,m)=-\lambda$).This 
can be easily seen from eqs.(44) and (45).
Since the thermal 
contribution to the renormalized coupling 
constant is negative 
someone could enquiry: is it possible for 
the renormalized coupling 
constant to vanish?  Once $\Delta\lambda(\beta)$ 
is $O(\lambda^{2})$ 
and we assume $D=4$, it is not possible  
to implement 
such a mechanism for finite temperatures. 
For $D<4$ the renormalized coupling constant 
is not necessarily a 
small quantity and it can even become a 
large quantity, due to its positive 
dimension $4-D$ in terms of the mass parameter 
$\mu$ (or using the 
language of critical phenomena, due to 
its positive dimension $4-D$ in terms 
of the 
scale $\frac{1}{a}$ where $a$ is 
the lattice spacing). Therefore 
we conclude that in the neighbourhood of $D=4$,
the renormalized coupling 
constant $\lambda(\beta)$
could vanish only for very high temperatures.
As we consider smaller spacetime 
dimensions the temperature where 
$\lambda(\beta)$ vanishes becomes 
lower and lower. 
For instance, for 
$D=3$ we expect to find a finite temperature 
$\beta^{-1}_{\star}$ 
such that the renormalized coupling constant 
vanishes.

We note 
that there is no discontinuity 
in the behavior between the cases $D=4$ and $D<4$ as we 
will see later (see 
eq.(49)). For $D<4$ the model becomes 
superrenormalizable and 
only a finite number  
set of graphs need overall counterterms. In the 
one-loop aproximation for $D=4$ there are only two divergent 
graphs and for $D<4$ there is only 
one. This result can be 
easily obtained by investigating eq.(38). In this equation the 
divergent part of the effective 
potential is given 
by $\Gamma(s-\frac{D}{2})$ 
and for $D<4$ only the $s=1$ pole 
will appear. In other words, for $D<4$ 
there is only finite coupling constant 
renormalization at the one-loop 
aproximation. The graph $s=2$ gives a 
finite and negative contribution 
to the coupling constant. For $D\geq 4$ the renormalization of 
the coupling constant is obligatory 
(note the presence of the pole in 
$s=2$). Going back to the $D$-dimensional case, the 
renormalization conditions also are 
given by eqs.(40) and (41).
Using the renormalization conditions 
in eq.(38), we can find  the regular 
part of the analytic extension which 
gives a finite contribution to the 
renormalized mass squared 
$\Delta m^{2}(D,m,\lambda,\beta)$ and coupling 
constant $\Delta\lambda(D,m,\lambda,\beta)$ in a $D$-dimensional 
flat spacetime. We will simplify 
the notation writing  $\Delta m^{2}(\beta)$ 
and $\Delta\lambda(\beta)$. For 
even $D$  they are given respectively  by:
\begin{equation}
\Delta m^{2}(\beta)=\frac{\mu^{D-2}
\lambda}{2(2\pi)^{D/2}}\biggl(\frac{(-1)^
{\frac{D}{2}-1}}{(\frac{D}{2}-1)!}
\psi(\frac{D}{2})(\frac{m}{\mu})^{D-2}+
\sum^{\infty}_{n=1}\biggl(\frac{m}
{\mu^{2}\beta n}\biggr)^{\frac{D}{2}-1}
K_{\frac{D}{2}-1}(mn\beta)\biggr)
\end{equation}
and
\begin{equation}
\Delta\lambda(\beta)=-\frac{3}{2}\frac{\mu^{D-4}\lambda^{2}}
{(2\pi)^{D/2}}\biggl(\frac{(-1)
^{\frac{D}{2}-2}}{(\frac{D}{2}-2)!}
\psi(\frac{D}{2}-1)(\frac{m}{\mu})^{D-4}+
\sum^{\infty}_{n=1}\biggl(\frac{m}
{\mu^{2}\beta n}\biggr)^
{\frac{D}{2}-2}K_{\frac{D}{2}-2}(mn\beta)\biggr)
\end{equation}
where $\psi(s)=\frac{d}{ds}ln\Gamma(s)$. 
For odd $D$, the first 
term between parentesis in eqs.(46) 
and (47) must be 
replaced by $\Gamma(1-\frac{D}{2})
(\frac{m}{\mu})^{D-2}$ and 
$\Gamma(2-\frac{D}{2})(\frac{m}{\mu})^{D-4}$ 
respectively.
 The first terms between parentesis 
of eq.(46) and eq.(47) are 
temperature independent therefore 
it is possible to isolate the thermal 
contribution to the renormalized 
mass and coupling constant in a 
generic $D$-dimensional spacetime 
in the one-loop aproximation. 
Using eq.(46) and eq.(47) we obtain the 
following contribution to the thermal mass 
and coupling constant respectively:
\begin{equation}
\Delta m^{2}(\beta)-\Delta m^{2}(\infty)=
\frac{\mu^{D-2}\lambda}{2(2\pi)^{D/2}}
\sum^{\infty}_{n=1}\biggl(\frac{m}{\mu^{2}\beta n}\biggr)^
{\frac{D}{2}-1}K_{\frac{D}{2}-1}(mn\beta)
\end{equation}
and
\begin{equation}
\Delta \lambda(\beta)-
\Delta\lambda(\infty)=-\frac{3}{2}\frac{\mu^{D-4}
\lambda^{2}}{(2\pi)^{D/2}}\sum^{\infty}_{n=1}
\biggl(\frac{m}{\mu^{2}\beta n}\biggr)^
{\frac{D}{2}-2}K_{\frac{D}{2}-2}(mn\beta).
\end{equation}

 These are among the main results of the paper. 
Since $\Delta\lambda(\beta)-
\Delta\lambda(\infty) <0$ we may 
have a temperature $\beta_{\star}^{-1}$ 
where $\lambda(\beta)$ vanish for 
$D<4$. Our result is different from the 
Frohlich result \cite{15} 
in which all all the Green's functions of the theory for 
$D>4$ correspond to a free field i.e. the 
model is gaussian at zero temperature above 
four spacetime dimensions. 
In our case, the higher $2n$-point functions are not zero 
as was discussed by Weldon 
\cite{43}.

Before discussing a existence of a first order phase transition, 
we would like to point out that the investigation of the 
$(\lambda\varphi^{4})_{D=4}$ model with a negative bare 
coupling constant has recently been done by Langfeld et al, 
where an analytic continuation of 
the model with positive $\lambda$ 
to negative values was presented \cite{18}.  Although several  
authors claim that the renormalized 
coupling constant of the $\lambda\varphi^{4}$ 
model must 
be positive, a definitive supporting argument 
is still lacking. Previous investigations 
have been done by many authors \cite{23}. 
We would like to stress that 
the sign of the renormalized coupling 
constant is not fixed by the 
renormalization procedure in the $(\lambda\varphi^{4})_{D=4}$. 
Gallavoti 
and Rivasseau discussed examples with 
positive bare coupling constant 
where different cutoffs lead to 
renormalized coupling constants with 
different signs \cite{24}.

Going back to the discussion of a first order 
phase transition, let us  
define a dimensionless effective 
potential $v=\frac{V}{\mu^{D}}$,
 as: 
\begin{eqnarray}
v(\beta,\phi) &=& \frac{1}{2}m^{2}\mu^{2-D}\phi^{2} + \frac{\lambda}{4(2\pi)^{D/2}}\sum^{\infty}_{n=1}\biggl
(\frac{m}{\mu^{2}\beta n}\biggr)^{\frac{D}{2}-1}K_{\frac{D}{2}-1}
(mn\beta)\phi^{2}\nonumber \\
& + & \frac{\lambda}{4!}\mu^{4-D}\phi^{4}
-\frac{1}{16}\frac{\lambda^{2}}{(2\pi)^{D/2}}
 \sum^{\infty}_{n=1}
\biggl(\frac{m}{\mu^{2}\beta n}\biggr)^
{\frac{D}{2}-2}K_{\frac{D}{2}-2}(mn\beta)\phi^{4}\nonumber\\
&+&\ high\ order\ terms\ in\ s.
\end{eqnarray}
In the effective potential all the powers $\phi^{2s}$ 
of the field will appear as stated in eq.(38). 
For instance, the term 
corresponding to the 
$2s-th$ power of the field is proportional to
\begin{equation}
\sum^{\infty}_{n=1}\biggl(\frac{m}{\mu^{2}\beta n}\biggr)^
{\frac{D}{2}-s}K_{\frac{D}{2}-s}(mn\beta)\phi^{2s}.
\end{equation}
The previous results can be used to demonstrate a 
first order phase transition in the 
$(\lambda\varphi^{4})_{D=3}$ model. 
To simplify our discussion 
let us assume that is 
possible to truncate the series of the effective potential in 
$s=3$. These does not imply the assumption 
that high order powers of the 
field gives vanishing contributions. 
They are simply neglected 
as compared to the leading terms, 
since we are interested in the profile 
of the effective potential 
near the origin. The coefficient 
of $\varphi^{6}$ is positive 
(one requires this to ensure that the truncated 
effective potential is bounded from below). 
For the sake of simplicity, 
let us also assume that the coefficient of 
the $\varphi^{6}$ is constant and given by 
$\sigma$ for both cases $D=3$ and 
$D=4$.
In these 
cases the leading contributions 
to the effective potential are respectively:
\begin{eqnarray} 
v(\beta,\phi) &=& \biggl(\frac{1}{2}m^{2}+
\frac{\lambda}{4(2\pi)^{\frac{3}{2}}}
\sum^{\infty}_{n=1}\biggl(\frac{m}{\mu^{2}\beta n}\biggr)^
{\frac{1}{2}}K_{\frac{1}{2}}(mn\beta)\biggr)
\phi^{2}\nonumber\\
&+&\biggl(\frac{\lambda}{4!}-
\frac{\lambda^{2}}{16(2\pi)^{\frac{3}{2}}}
\sum^{\infty}_{n=1}\biggl(\frac{m}{\mu^{2}\beta n}\biggr)^
{-\frac{1}{2}}K_{\frac{1}{2}}(mn\beta)\biggr)
\phi^{4}+\sigma\phi^{6},
\end{eqnarray}
and
\begin{eqnarray}
v(\beta,\phi)&=&\biggl(\frac{1}{2}m^{2}+
\frac{\lambda}{16\pi^{2}}
\sum^{\infty}_{n=1}\biggl(\frac{m}{\mu^{2}
\beta n}\biggr)K_{1}(mn\beta)\biggr)\phi^{2}\nonumber\\
&+&\biggl(\frac{\lambda}{4!}-\frac{\lambda^{2}}{64\pi^{2}}
\sum^{\infty}_{n=1}K_{0}(mn\beta)\biggr)
\phi^{4}+\sigma\phi^{6}.
\end{eqnarray} 
From the above discussion, for $D<4$ we obtain 
the following profile for the 
effective potential in the neighborhood of the 
origin.  
Bellow the temperature $\beta^{-1}_{\star}$, 
the dimensionless 
effective potential has only one global minimum. 
Heating the system 
above the temperature $\beta^{-1}_{\star}$, the renormalized 
coupling constant would become negative and the 
system can develop 
a first order phase transition since the 
expectation value of the order parameter 
changes discontinuously by temperature effects.
The situation is similar to the Coleman-Weinberg 
mechanism for massless fields. The effects of 
the quantum corrections is towards the 
direction of breaking a symmetry. Note 
the similarity with the tricritical phenomena 
where in the tree level 
$(V(\varphi)=m^{2}\varphi^{2}+\lambda\varphi^{4}+
\sigma\varphi^{6})$  the model 
develop a first order phase transition if we allow the 
coefficient of the quartic term to be negative
\cite{48}.

In a detailed study, using the ring-improved one-loop effective 
potential, Arnold and Spinosa \cite{50} showed 
that even for temperature independent coupling constant, the 
$\lambda\varphi^{4}$ model can develop 
at the first sight a first order phase transition.
Nevertheless, these authors verified that 
the contribution of 
higher loop corrections dominates over the 
one-loop ring improved 
contributions. By these reasons, in this 
approximation they cannot distinguish 
between a first or a second order phase transition.
As we discussed in the introduction,
the thermal correction to the 
coupling constant if we include high order loops 
in the effective potential is positive for 
high temperatures. 
Nevertheless for low temperatures 
the effective renormalized coupling 
constant may become negative. In this case we 
still have a first order phase transition.
From the above discussion, 
we have obtained the following 
result: in the massive $\lambda\varphi^{4}$ for $D<4$ 
for temperatures above $\beta^{-1}_{\star}$ the 
effective potential will develops a local minimum 
at the origin (a false vacuum) and a global one outside 
the origin. In this case
the initial 
metastable phase may decay to 
a stable one by nucleation of bubbles. 
The temperature is the parameter that drives 
the first order phase 
transition. Evaluating the ring diagrams Carrington  and 
Takahashi independently obtained 
in a pure scalar model at $D=4$ results 
which are consonant with ours results in $D=3$ \cite{33}.

\section{ The one-loop effective potential in the massless 
Gross-Neveu model at finite temperature.}\

Our purpose throughout this section is to examinate 
the behavior of the renormalized coupling constant in a 
model involving fermions with a quartic interaction.
In two-dimensional 
spacetime $(D=2)$ the 
model is renormalizable and  ultraviolet asymptotically free. 
We will consider an N-component fermion 
field where the limit of 
large N will be investigated. As it was 
discussed in ref.(4), due to the quartic nature
 of the interaction, it is possible to 
introduce an ultralocal 
auxiliar scalar field $\varphi$ which is 
formally equal to 
$g\overline{\psi}\psi$ where $\psi(x)$ is the fermionic field, 
in order to present the effective potential of the model. 
As we did in section II, we suppose that 
the quantum field is in 
thermal equilibrium with a reservoir at 
temperature $\beta^{-1}$. 
We will show that for $D=2$ and $D=4$ inthe one-loop 
approximation the 
sign of the thermal 
correction to the renormalized coupling 
constant cannot be 
calculated. On the other hand, for $D=3$ in the 
one-loop aproximation the thermal correction to the 
renormalized coupling constant is zero.

The Lagrange density of the massless model is given by:
 
\begin{equation}
{\cal L}(\overline{\psi},\psi, \varphi)=i\overline{\psi}
\gamma^{\mu}\partial_{\mu}\psi-\frac{1}{2}\varphi^{2}-
g\varphi\overline{\psi}\psi.
\end{equation}
Defining $\varphi_{0}$ as the vacuum expectation value of 
$\varphi$, i.e. $\varphi_{0}=<0|\varphi|0>= 
<0|g\overline{\psi}\psi|0>$, 
the leading terms in the effective potential 
for large N are given by the 
tree-level graphs plus all one-loop graphs,
\begin{equation}
V(\varphi)=\frac{1}{2}\varphi_{0}^{2}
-iN\sum^{\infty}_{s=1}\frac{1}{2s}(g\varphi_{0})^{2s}
\int\frac{d^{D}q}{(2\pi)^{D}}\frac{1}{k^{2s}}.
\end{equation}

After a Wick rotation we identify the effective 
potential as the free energy of the system. At zero 
temperature the model has a spontaneous breakdown of the 
chiral symmmetry where the fermions acquire mass. The 
symmetry is restored at finite temperature by a second order 
phase transition \cite{2}. This result can 
be obtained by summing the series in the effective potential.
Since we are interested only in the thermal behavior 
of the mass and coupling constant instead of repeat the 
well known calculations we will adopt a very unusual road, 
similar to the previous chapter, by regularizing each 
term of the series in the effective potential before summing up.

To introduce finite temperature effects we 
assume that the Grassmannian 
integration in the path integral goes over 
anti-periodic configurations in 
Euclidean time. In the effective potential 
this is equivalent to the  
replacement given by eq.(21) and
\begin{equation}
\omega\rightarrow\frac{2\pi}{\beta}(n+\frac{1}{2}).
\end{equation}
Using eq.(33) and defining $f(D,s)$ by: 
\begin{equation}
p(D,s)=\frac{1}{2^{2s+1}}\frac{1}{\pi^{2s-
\frac{d}{2}}}\frac{(-1)
^{s}}{s}\frac{\Gamma(s-\frac{d}{2})}{\Gamma(s)},
\end{equation}
it is not difficult to show that 
$V(\beta,\varphi_{0})$ is given by:
\begin{equation}
V(\beta,\varphi_{0})= \frac{1}{2}\varphi_{0}^{2}+
N\sum^{\infty}_{s=1}p(D,s)(g\varphi_{0})
^{2s}\beta^{2s-D}
\sum^{\infty}_{n=-\infty}\frac{1}{(n+\frac{1}{2})^{2s-d}}.
\end{equation}
 Note that we are using 
dimensional regularization in 
eq.(55) and it is well known that 
for massless fields this technique 
requires modification in 
order to deal with infrared 
divergences \cite{29}. Since we 
are regularizing only a 
$d=D-1$ dimensional integral, this 
procedure is equivalent 
to inserting a mass into the $d$ dimensional integral.
In other words, the Matsubara frequency plays the role of a 
"mass" in the integral, provided we exclude the limit 
$\beta\rightarrow\infty$, which means that we must 
restrict ourselves to non-zero temperature.   
 
Again, as in eq.(30), we can define a new field $\phi=
\frac{\varphi_{0}}{\mu}$ (no confusion must be done between 
the present auxiliar scalar field and the previous scalar field). 
Using eq.(24) we obtain
\begin{equation}
V(\beta,\phi)=\frac{1}{2}\mu^{2}\phi^{2}+
N\mu^{D}\sum^{\infty}_{s=1}
p(D,s)a^{\frac{D}{2}-s}(g\phi)^{2s}\sum^{\infty}_{n=
-\infty}\frac{1}{(n+\frac{1}{2})^{2s-d}}.
\end{equation}
The Hurwitz zeta function is defined as
\begin{equation}
\zeta(z,q)=\sum^{\infty}_{n=0}\frac{1}{(n+q)^{z}}
\end{equation}
for $Re(z)>1$ and $q\neq{0,-1,...}.$
 For $q=1$ we recover the usual Riemann zeta function. 
Defining:
\begin{equation}
r(D,s)=p(D,s)\biggl(\zeta(2s-d,\frac{1}{2})+(-1)^{2s-d}
\zeta(2s-d,-\frac{1}{2})-\frac{1}{2^{d-2s}}\biggr)
\end{equation}
the effective potential can be written as:
\begin{equation}
V(\beta,\phi)=\frac{1}{2}\mu^{2}\phi^{2}+N\mu^{D}\sum^
{\infty}_{s=1}r(D,s)a^{\frac{D}{2}-s}(g\phi)^{2s}.
\end{equation}
The effective potential is still baddly defined and it will 
be regularized by the principle of analytic extension. The 
function $r(D,s)$ is valid in the begining in an open 
connected set of points, i.e. for $Re(z)>1$. Since we are 
considering even non perturbative 
renormalizable models, let us study 
the cases $D=2,3$ and $4$. We would like 
to stress that even 
for the non perturbative renormalizable models 
it is possible to make qualitative predictions 
and we will 
regularize and renormalize the model in the standard way. 
A strong argument in favor of the study 
of the Gross-Neveu 
model is that the non-renormalizability does not appear in 
the leading $\frac{1}{N}$ approximation for $D=3$.

After the analytic continuation, the effective potential 
requires a renormalization procedure in the points $s=1,2..$ 
The renormalization condition which will fix the form of the 
counterterm  of the pole $s=1$ is:
\begin{equation}
\frac{\partial^{2}V}{\partial\phi^{2}}|_{\phi=cte}=\mu^{2}
\end{equation}
Due to infrared divergences, we must follow 
Coleman and Weinberg 
\cite{9} and choose the renormalization point at non-zero $\phi$. 
In order to evaluate the renormalized effective potential it is
necessary to use the Hermite formula of 
the analytic extension 
for the Hurwitz zeta function given by \cite {16}
\begin{equation}
\zeta(z,q)=\frac{1}{2q^{z}}+
\frac{q^{1-z}}{z-1}+2\int^{\infty}_{0}
(q^{2}+y^{2})^{\frac{-z}{2}}
\sin(z\arctan\frac{y}{q})\frac{1}{e^{2\pi y}-1}dy.
\end{equation}
 It is not difficult to show that 
the thermal contribution to 
the renormalized coupling constant is,
\begin{equation}
\Delta g(\beta)=N\mu^{D-2}\sum^{\infty}_{s=1}
r(D,s)(2s)
(2s-1)g^{2s}(\beta\mu)^{2s-D},
\end{equation}
 where it is understood that the polar terms 
in the summation have 
been subtracted remaining just the 
regular part of the analytic continuation.
 The situation is different from the 
massive $\lambda\varphi^{4}$ model, 
since we have the contribution of all 
terms of the series in $s$ and the sign 
of the thermal contribution to the 
renormalized coupling constant 
cannot be easily obtained. Nevertheless, for
sufficiently small $g$ the leading term is $O(g^{2})$. In this case, 
for $D=3$ and using the fact 
that $\zeta(0,q)=\frac{1}{2}-q$, 
we obtain that $\Delta g=0$. 
We found here that there is no 
thermal correction to the coupling constant at least in the 
one-loop approximation.
Note that $\Delta 
g(\beta)$ is still not well behaved. 
The terms $s>\frac{D}{2}$ 
are divergent in the low temperature limit (the use of dimensional 
regularization in the begining of the 
calculations leads to this situation). 
For $s<\frac{D}{2}$, the high temperature 
limit of the model is 
problematic due to the well known fact 
that ultraviolet divergences 
are worst as the spacetime dimension increases.

\section{Conclusions}

In this paper we studied the renormalization 
program assuming that scalar 
or fermionic fields are in equilibrium with 
a thermal reservoir at 
temperature $\beta^{-1}$. We have attempted 
to analize the consequences of 
the fact that not only the renormalized mass, but also 
the renormalized coupling constant 
acquire thermal corrections.

It is well known that if we have a one spatial 
dimension compactified system at a finite temperature, which has a 
spontaneous symmetry breaking there are two different 
ways to restore the symmetry.
Since the compactification of one spatial 
dimension gives us the well 
known mechanism of topological generation 
of mass, it is possible 
to restore the symmetry by thermal or 
topological effect. There is a very 
simple way to interpret the origin of 
the thermal and topological mass 
and coupling constant. The 
effective potential is not well defined. 
Using the Principle of the Analytic 
Extension, we regularize the model and 
the introduction of counterterms 
remove the principal part of the analytic extension, 
and the model becomes 
finite. Meanwhile, in the neighbourhood 
of the poles, the regular part of the 
analytic extension does not vanish. 
These temperature 
dependent regular part 
around the poles $s=1$ and $s=2$ (for $D=4$) are identified with 
the thermal correction to 
the mass and coupling constant.

It 
was proved that in the $\lambda\varphi^{4}$ model, 
in the one-loop aproximation, the 
thermal correction to the renormalized 
mass is positive and 
the thermal correction to the renormalized 
coupling constant is negative. 
In this case 
the renormalized coupling constant 
attains its maximum at zero temperature and 
decreases monotonically as the temperature 
increases. Since in $D=4$, 
$\Delta\lambda(\beta)$ is $O(\lambda^{2})$ 
it is not possible to vanish the 
renormalized coupling constant at a finite 
temperature of the thermal 
bath.  For strong couplings    
($D<4$) there is a finite   
temperature where this can be achieved.
For temperatures $\beta^{-1}>\beta^{-1}_{\star}$
(negative coupling 
constant) the system can develop a first order 
phase transition, where 
the origin is a false vacuum.

It is not all clear for us if at $D=4$ the system can
develop a first order phase transition. We are using the 
following argument to disregard such possibility. As we 
discussed in the introduction, in the two-loops approximation 
at high temperatures the thermal correction to the 
coupling constant is positive. The fact that in $D=4$ the 
model has a small zero temperature 
coupling constant eliminate the 
first order phase transition in $D=4$.

We would like to emphasize that the massive $\lambda\varphi^{4}$ 
model does not belong to the same universality class of 
the Ising model.
It is well known that it is possible to 
compare the $\lambda\varphi^{4}$ model in continuous 
$D$-dimensional Euclidean space with the Ising model.  One 
lattice formulation can be done and the 
continuum limit of the model ($a\rightarrow 0$, where $a$ is the 
lattice spacing) exist if the correlation length goes to infinite.
This fact implies that at the continuum limit of the 
lattice model the system must suffer a second order phase 
transition. In other words, close to the 
critical temperature a D-dimensional Ising model has the same 
correlation functions as those 
for a field theory ($\lambda\varphi^{4}$ model) defined in a 
D-dimensional Euclidean space near the critical 
temperature. Since in the 
paper we assume that the tree level mass squared is always positive 
and we found that the thermal mass squared is also positive, 
we are always far from the 
critical temperature. By these reasons the system cannot fall into 
the universality class of a Ising model.

 The analysis of this paper suggest two 
possible directions. First, we have to study 
the decay of the metastable 
ground state in the $(\lambda\varphi^{4})_{D<4}$ model 
evaluating the nucleation rate per unit 
volume in the system. The theory of bubbles nucleation at zero 
and finite temperature was proposed and developed 
by many authors \cite{47}. The basic result is that the 
probability per unit volume per unit time of the metastable vacuum 
to decay is given by $\Gamma=A e^{-S(\varphi)}$, where $S(\varphi)$ is 
the Euclidean action of the "bounce" solution which describes tunneling 
between a metastable and a true vacuum.
Another possible direction is to examinate if the 
metastability of the system  
(the false ground state) can be eliminated in a 
more general scalar model. This former subject will 
be presented soon in a 
forthcomming paper\cite{36}. We conclude the paper 
with some some questions which remain to be 
answered.

(i) Is the existence of the first order phase transition in 
$(\lambda\varphi^{4})_{D=3}$ an 
artifact of our approximation? It will be interesting 
to obtain a non-perturbative argument to demonstrate or disprove 
this fact in a general way. 

(i) Is the series given by eq.(69) Borel summable ? 
It is well known that 
the lack of Borel summability means that the system is unstable, since the 
vacuum to vacuum amplitude develop and imaginary 
part \cite{25}. It would be 
interesting to investigate these questions.

\section{Acknowlegement}

We would like to thanks 
Prof.L.H.Ford, Prof.A.Grib and Prof.C.A.Carvalho for useful 
comments and criticisms and for valuable discussions. We are also 
greateful to Dr.T.Vachaspati and Prof.A.Vilenkin
 for many helpful discussions. N.F.Svaiter 
would like to thanks the
hospitality of the Institute of Cosmology, Tufts University,
where part of this
work was carried out. This paper was
suported by Conselho Nacional de 
Desenvolvimento Cientifico e Tecnologico do Brazil (CNPq).

\end {document}